\documentclass[apjl]{emulateapj}
\usepackage{graphics}
\usepackage{enumitem}
\usepackage{amsmath}
\usepackage{xcolor}
\usepackage{soul}

\newcommand{\chandra}{\textit{Chandra }}

\begin{document}


\shortauthors{TEMIM ET AL.}

\shorttitle{High Velocity Pulsar in MSH 15-5\textit{6}}

\title{Proper Motion of the High-Velocity Pulsar in SNR MSH 15-5\textit{6}}


\author{Tea Temim\altaffilmark{1},Patrick Slane\altaffilmark{2}, Paul P. Plucinsky\altaffilmark{2}, Joseph Gelfand\altaffilmark{3}, Daniel Castro\altaffilmark{2}, and Christopher Kolb\altaffilmark{4}}


\altaffiltext{1}{Space Telescope Science Institute, 3700 San Martin Drive, Baltimore, MD 21218, USA}
\altaffiltext{2}{Harvard-Smithsonian Center for Astrophysics, 60 Garden Street, Cambridge, MA 02138, USA}
\altaffiltext{3}{New York University, Abu Dhabi, United Arab Emirates}
\altaffiltext{4}{North Carolina State University}

\slugcomment{Accepted by ApJ}

\begin{abstract}

We present a measurement of the proper motion of the presumed pulsar in the evolved composite supernova remnant (SNR) MSH~15-5\textit{6} whose pulsar wind nebula (PWN) has been disrupted by the supernova (SN) reverse shock. Using \textit{Chandra} X-ray observations acquired over a baseline of 15~years, we measure a pulsar velocity of $720^{+290}_{-215}\rm \: km\: s^{-1}$ and a direction of motion of $14^{\circ}\pm22^{\circ}$ west of south. We use this measurement to constrain a hydrodynamical model for the evolution of this system and find that its morphology is well-described by an SNR expanding in an ambient density gradient that increases from east to west. The effect of the density gradient and the pulsar's motion is an asymmetric interaction between the SN reverse shock and the PWN that displaces the bulk of the PWN material away from the pulsar, towards the northeast. The simulation is consistent with an SNR age of 11,000~years, an SN ejecta mass of 10~$\rm M_{\odot}$, and an average surrounding density of $0.4\rm \: cm^{-3}$. However, a combination of a higher SN ejecta mass and ambient density can produce a similar SNR morphology at a later age.

\end{abstract}

\section{Introduction} \label{intro}

Composite supernova remnants (SNRs) are characterized by an SNR shell produced by a supernova (SN) blast wave and a pulsar wind nebula (PWN) generated by a rapidly-rotating, highly magnetized neutron star. The evolution of these two components is closely coupled. The PWN initially drives a shock into the innermost SN ejecta and eventually encounters the SN reverse shock that propagates towards the explosion center \citep[e.g.][]{blondin01}. The signatures of the interaction that follows provide insight into the evolution of the system, as well as constraints on the SN progenitor, density structure of the ambient ISM, and the evolution of the spectrum of the relativistic particle population injected by the pulsar that ultimately escapes into the ISM.

Composite SNRs in which the reverse shock has already begun interacting with the PWN exhibit complex morphologies consisting of irregularly-shaped PWNe that are displaced from the geometric centers of their SNRs \citep[e.g.][]{temim09,slane12}. Recent multi-wavelength studies and hydrodynamical (HD) modeling of these evolved systems have significantly increased our understanding of how these morphologies form \citep{temim15,kolb17}. The structure of the composite SNR G327.1-1.1 was successfully reproduced by an HD simulation invoking a density gradient in the surrounding ISM and a rapidly moving pulsar that effectively cause the reverse shock to interact with the PWN asymmetrically. \citet{kolb17} used the same HD model to explore what parameters govern the evolution of composite SNRs that contain PWNe generated by pulsars with high kick velocities. They found that the structure of these SNRs is most strongly influenced by the structure of the ambient medium and the total spin-down energy of the pulsar. Secondary factors include the pulsar's velocity, direction of motion, and its initial spin down luminosity. The simulations also show that the passage of the reverse shock leaves a trail of PWN material behind the pulsar and that the direction of this trail reflects the flow of the SN ejecta at that location. Since a number of factors drive the morphology of composite SNRs, observational constraints on some of these input parameters are necessary to produce a realistic model and learn about other SNR properties. In the case of SNR G327.1-1.1 \citep{temim15}, the radio and X-ray observations provided constraints on the pulsar's spin-down luminosity and the orientation of the ISM density gradient, which in turn allowed us to infer the direction of the pulsar's motion from the HD simulation. Here, we apply the HD model to another evolved composite SNR, MSH~15-5\textit{6}.

\begin{deluxetable*}{crrrrrr}
\tablecolumns{7} \tablewidth{0pc} \tablecaption{\label{tab1}POINT SOURCE OFFSETS}
\tablehead{
\colhead{Source} & \multicolumn{4}{c}{Coordinates (J2000)} & \multicolumn{2}{c}{Offset} \\
\colhead{}  & \multicolumn{2}{c}{2001} & \multicolumn{2}{c}{2016} & \multicolumn{2}{c}{(arcsec)} \\
\colhead{} & \colhead{RA (deg)} & \colhead{Dec (deg)} & \colhead{RA (deg)} & \colhead{Dec (deg)} & \colhead{$\Delta$RA} & \colhead{$\Delta$Dec}
}
\startdata
1 & 238.08776 & -56.34078 & 238.08774 & -56.34082 & 0.096 & 0.153 \\
2 & 238.05869 & -56.34383 & 238.05878 & -56.34377 & -0.305 & -0.218 \\
3 & 238.13282 & -56.36208 & 238.13274 & -56.36209 & 0.300 & 0.028 \\
4 & 238.11162 & -56.28458 & 238.11164 & -56.28459 & -0.028 & 0.036 \\
Pulsar & 238.05275 & -56.31615 & 238.05271 & -56.31630 & 0.134 & 0.532
\enddata
\end{deluxetable*}

\begin{figure*}
\epsscale{1.05} \plotone{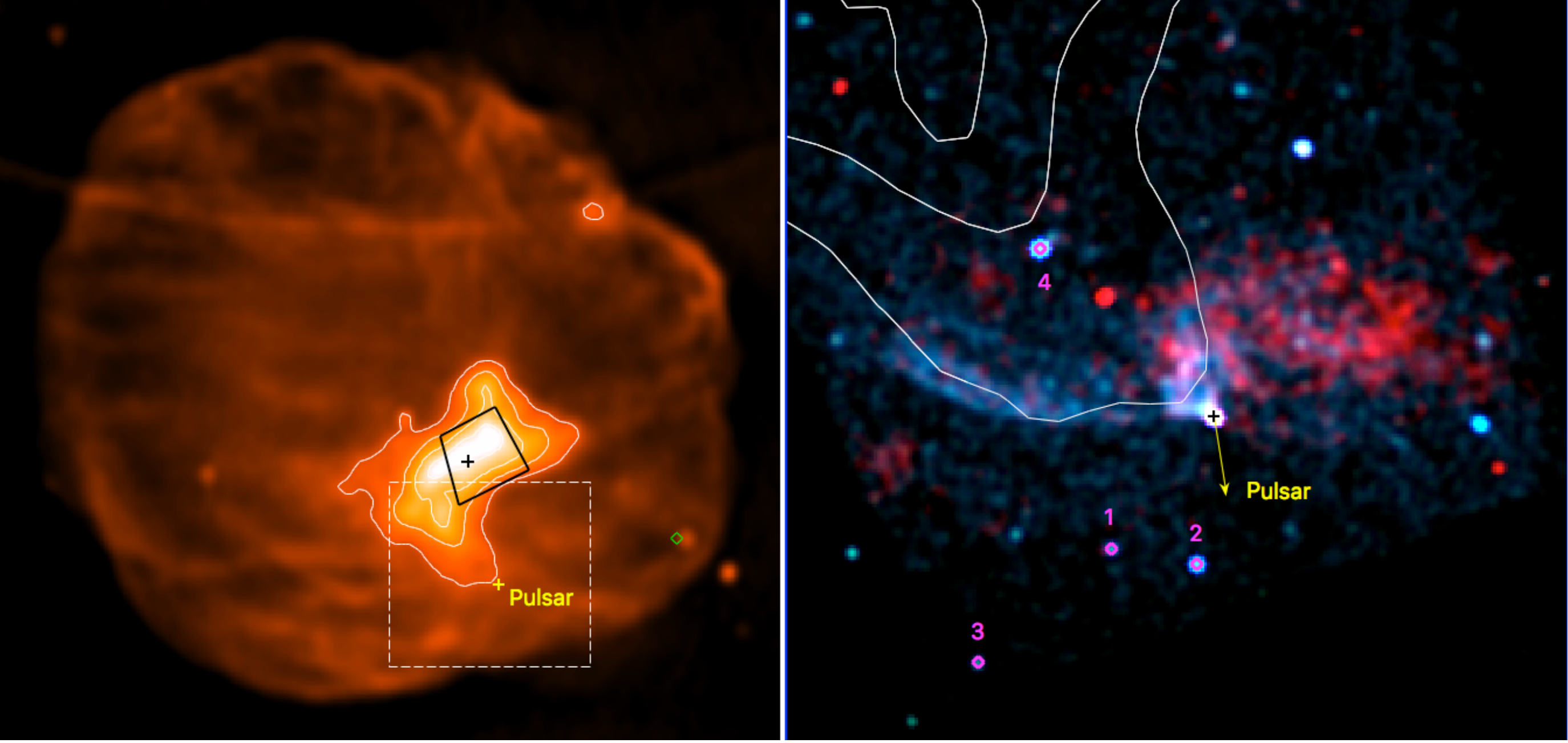}\caption{\label{fig1}Left: MOST 843 MHz radio image of MSH 15-5\textit{6} \citep{whiteoak96} showing the SNR shell and a PWN outlined by the white contours. The location of the pulsar is indicated by the yellow cross. The black cross and represents the SN explosion center calculated from the pulsar's proper motion, assuming a distance of 4.1 kpc and an SNR age of 11,000 years. The surrounding black box represents the uncertainty in the explosion center derived from the uncertainty in the pulsar's velocity and direction of motion. The region in the white dashed square is shown in the right panel. Right: The full 175 ks \textit{Chandra} X-ray image of the pulsar and PWN in MSH 15-5\textit{6} with the 0.3-1.5 keV emission shown in red and 1.5-9.0 keV emission in blue. The location of the pulsar is indicated by the black cross and the MOST radio contours are overlaid in white. The yellow arrow represents the pulsar's direction of motion, as measured from the pulsar's offset between the 2001 and 2015 observations. The sources used to register the images are marked by magenta circles and their coordinates and offsets are listed in Table \ref{tab1}.}
\end{figure*}

The radio and X-ray morphologies of MSH~15-5\textit{6} are shown in Figure~\ref{fig1}. The 843 MHz radio image from the Molonglo Observatory Synthesis Telescope (MOST) \citep{whiteoak96} is shown in the left panel of the figure and consists of filamentary, circular SNR shell, approximately 18\arcmin\ in radius. The PWN is outlined by the white contours and is offset to the southwest of the geometric center of the SNR shell. Its radio luminosity of $L_{10^7-10^{11}Hz}=5\times10^{34}\:\rm erg\:s^{-1}$ is the third highest PWN luminosity after the Crab Nebula and SNR G328.4+0.2 \citep{dickel00}, assuming a distance of 4.1 kpc obtained from H$\alpha$ velocity measurements \citep{rosado96}. 
The presumed pulsar is located at the tip of the radio PWN, close to the southwestern edge of the SNR shell. Its position is marked by the yellow cross in the left panel of Figure~\ref{fig1}. The pulsations from the source have not been detected, but the presence of the pulsar is inferred from the surrounding PWN. The location and structure of the displaced PWN itself suggest that the SNR reverse shock has already interacted with the nebula. Based on the spectral properties of the thermal X-ray emission in the SNR shell attributed to swept-up interstellar medium (ISM) material, \citet{temim13a} estimated an age of 16,500 years, consistent with the SNR being an evolved system. In this paper, we measure the proper motion of the presumed pulsar in MSH~15-5\textit{6} and use it to constrain an HD model of this system in order to understand the SNR's evolution, ambient environment, and the progenitor star.

\begin{deluxetable}{llc}
\tablecolumns{3} \tablewidth{20pc} \tablecaption{\label{tab2}HD MODEL PARAMETERS} 
\tablehead{\colhead{Parameter} & \colhead{Description} & \colhead{Value}} 
\startdata
\\
\underline{Input Parameters:} & & \\

*$v_p$ (km/s) & Pulsar velocity & 720 \\
*$v_p$ direction & West of South & 10$^{\circ}$ \\
$E_{51}$ ($\rm 10^{51}\:erg$) & Explosion energy & 1.0 \\
$M_{ej}$ ($\rm M_{\odot}$) & SN ejecta mass & 10 \\
$n$ & SN ejecta density profile index & 12 \\
$\dot{E_0}$ (erg/s) & Initial spin-down luminosity & $7.0\times10^{38}$ \\
$b$ & Pulsar braking index & 3.0 \\
$\tau_0$ (yr) & Spin-down timescale & 2000 \\
*$n_0$ ($\rm cm^{-3}$) & Minimum ambient density & 0.1 \\
$x$ & $n_0$ contrast of 8.1 & 1.125 \\
$H$ (pc) & $n_0$ characteristic length scale &  3.24 \\
$n_0$ gradient angle & North of West & 10$^{\circ}$ \\
\\
\underline{Simulation Output:} & & \\

*$R_{SNR}$ (pc) & SNR radius & 21 \\
*$L_{X(2-10)}$ (erg/s) & PWN X-ray luminosity & $4.0\times10^{34}$ \\
*$\dot{E}$ (erg/s) & Current spin-down luminosity & $1.7\times10^{37}$ \\
$t$ (yr)  & SNR age & 11,000 \\

\enddata
\tablecomments{The pulsar velocity $v_p$ is calculated from the pulsar proper motion measured in this work. The ejecta mass $M_{ej}$, braking index of the pulsar, initial spin-down luminosity, the spin-down timescale, and the SNR age were varied until the simulation produce the observed PWN morphology and an SNR radius of 21 pc, for an assumed distance of 4.1 kpc. The input and output properties marked by an asterisk are those that have been constrained by observations. The equation for the density gradient of the medium into which the SN expands is given in \citet{blondin01}.}
\end{deluxetable}



\section{Observations and Data Reduction} \label{obsv}

Our analysis includes \chandra X-ray observations of MSH 15-5\textit{6} carried out at two different epochs. The first observation was taken on 2001 August 18 for a total exposure time of 56.8 ks (observations ID 1965, PI: Plucinsky). The second set of five observations was carried out in 2016 May 30-June 4, for a total exposure time of 175 ks (ObsIDs 17897, 18862-18865, PI: Temim). All observations were acquired using the Advanced CCD imaging Spectrometer (ACIS-S) in the very faint (VFAINT) mode. We performed the standard cleaning and data reduction on both sets of observations using CIAO version 4.9 and calibration database version CALDB 4.7.3.

\section{Proper Motion of the Pulsar} \label{proper}

In order to search for the pulsar proper motion, we compared the 2001 \chandra image with the longest of the five observations from 2016 (ObsID 17897), carried out on 31 May for 78~ks. This provided a baseline of 14.78 years. The other four observations from 2016 were not used in the analysis, since they all had exposure times shorter than the 2001 observation. Since the detection sensitivity for reference sources to be used in aligning the two images is set by the shorter 2001 observation, it was not necessary to merge the 2016 images and use all 175~ks, especially since merging images taken at slightly different roll angles may introduce additional astrometric uncertainties. 

While the pointing accuracy of \chandra is $\sim$  0.4\arcsec, the relative astrometry between images can be improved by cross-matching sources in the two observations. The CIAO tool \textit{reproject\_aspect}, a script that runs the tools \textit{wcs\_match} and \textit{wcs\_update}, can be used to calculate astrometric shifts between two source lists and then apply the offsets to the observations. We first used the CIAO command \textit{wavdetect} to search for point sources in both images. Eighteen sources were identified in the 2016 image. We removed the pulsar from this source list, as well as six additional sources that we identified in VizieR to have proper motions of more than 6 $\rm mas\:yr^{-1}$. The remaining eleven sources were cross-matched with the source list from the 2001 observation using \textit{reproject\_aspect}. The script deleted bad matches that primarily consisted of far off-axis and blended sources, resulting in four matches that were used to calculate the correction to the aspect solution. These sources are shown in Figure~\ref{fig1}. We then ran \textit{wavdetect} on the corrected images and calculated the difference in RA and Dec for the four cross-matched sources and the pulsar. The final coordinates of the sources and their residuals are listed in Table~\ref{tab1}. The shifts in RA and Dec measured for the pulsar are 0.134$\arcsec$ and 0.532$\arcsec$, respectively. The average absolute values of $\Delta$RA and $\Delta$Dec for the cross-matched sources are 0.182$\arcsec$ and 0.134$\arcsec$, while the average uncertainty on the source positions measured by \textit{wavdetect} is 0.115$\arcsec$ in RA and 0.077$\arcsec$ in Dec. These two sets of uncertainties were combined to derive the final  uncertainty estimate in the residuals measured for the pulsar. Assuming a distance of 4.1 kpc, this translates to a pulsar velocity of $\rm 720^{+290}_{-215}\: km\: s^{-1}$ and a direction of motion of $14^{\circ}\pm22^{\circ}$ west of south (yellow arrow in the right panel of Figure~\ref{fig1}). This velocity is on the high end of the distribution of kick velocities estimated for other neutron stars inside SNRs \citep[e.g.][]{hollandashford17}.


\section{Observed Properties of the PWN}

Figure~\ref{fig1} (right) shows the 2016, 175~ks \textit{Chandra} X-ray image of the region surrounding the pulsar, with the 0.3-1.5 keV emission in red and 1.5-9.0 keV emission in blue. The pulsar is surrounded by an extended, but compact X-ray PWN. \citet{temim13a} analyzed the \textit{XMM-Newton} and 2001 \textit{Chandra} data of the PWN region, and found the total X-ray luminosity of the non-thermal emission to be $L_{X}(0.3-10\: keV)=4\times10^{34}\:\rm erg\:s^{-1}$ and the photon index of the compact PWN to be $1.84^{+0.13}_{-0.11}$. The soft X-ray emission to the west that is outside of the radio PWN contours has a significant thermal component that may arise from SN ejecta that have been encountered by the reverse shock. Thermal emission with enhanced abundances was also present throughout the region surrounding the PWN. An arc of hard X-ray emission traces the southern edge of the radio PWN. While the spectrum of the arc showed that both thermal emission with enhanced abundances and non-thermal emission are present in the region \citep{temim13a}, the hard non-thermal component is more prominent along the arc than in the western region that appears red in Figure~\ref{fig1}(right).
The nature of this arc is still unknown, but one possibility is that its emission originates from a pulsar's jet that has been swept back by the SN reverse shock. In this case, the thermal ejecta emission arises from projection along the line of sight.

\begin{figure*}
\center
\epsscale{1.1} \plotone{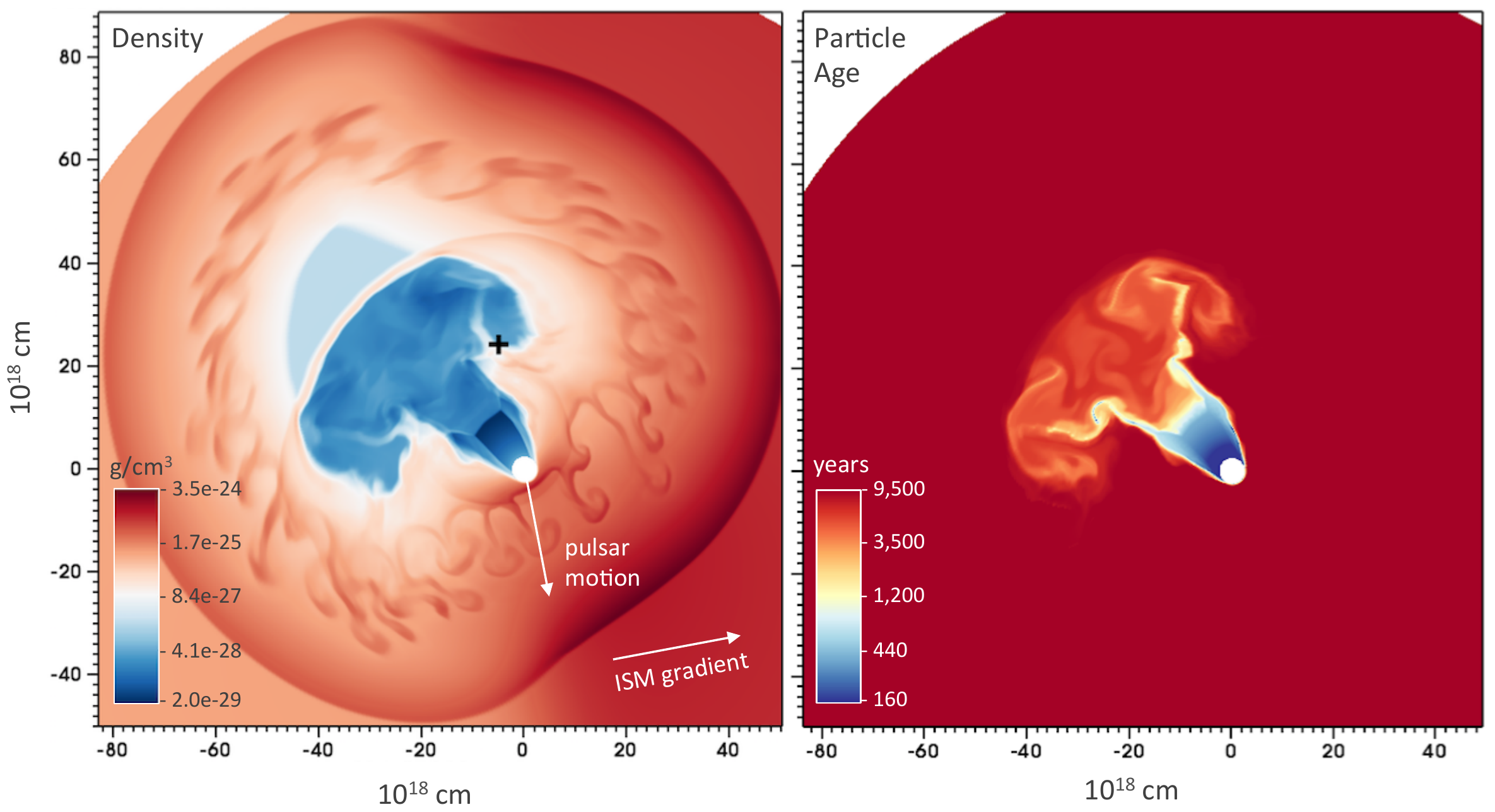}\caption{\label{fig2} HD simulation of an SNR expanding into an ISM with a density gradient, and hosting a PWN produced by a moving pulsar. The left panel shows a density map ($\rm g \: cm^{-3}$). The ISM density increases from east to west and the pulsar has a velocity of 720 $\rm km\: s^{-1}$ in a direction 10 degrees west of south, as measured in the work. The center of the of the explosion is marked by a black cross. The SNR morphology produced by the HD simulation at 11,000 years resembles the observed morphology of MSH~15-5\textit{6} (Figure~\ref{fig1}). The right panel shows the age of the pulsar-injected particles at an SNR age of 11,000 years for the same simulation as shown in the left panel. The younger particles shown in blue are expected to give rise to X-ray emission, while the aged particles shown in red would primarily emit in the radio. This is consistent with the X-ray and radio morphologies of MSH~15-5\textit{6} shown in Figure~\ref{fig1}.}
\end{figure*}

\section{Hydrodynamical Modeling}

We modeled the evolution of MSH~15-5\textit{6} using the VH-1 hydrodynamics code \citep{blondin01} that has been modified to simulate a composite SNR evolving into a non-uniform ISM and hosting a PWN generated by a moving pulsar. The modified code is discussed in detail in \citet{kolb17}. The PWN is modeled as an expanding bubble of relativistic gas (adiabatic index $\gamma$=4/3) driven by the pulsar's time-dependent spin-down luminosity $\dot{E}=\dot{E_0}(1+t/\tau_0)^{-(b+1)/(b-1)}$, where $\dot{E_0}$ is the pulsar's initial spin-down luminosity, $\tau_0$ is the spin-down timescale, and $b$ is the braking index. The braking index and the magnetic field are both assumed to be constant \citep[see][]{blondin01}. The SNR shell is modeled as a self-similar wave that expands into an SN ejecta density profile that falls off exponentially with a power-law index of $n=12$ \citep[e.g.][]{chevalier82}, a value that is typical for a red supergiant progenitor \citep{matzner99}. The surrounding ISM density profile includes a gradient of the form $n(z)=n_0[1-(2-x)tanh(z/H)/x]$, where $z$ is the distance from the center for the explosion, $n_0$ is the average (central) density, $H$ is the characteristic length scale over which the density varies, and $x$ is the parameter that defines the density contrast \citep{blondin01}. The simulation is designed in the rest frame of the pulsar and the pulsar's motion through the SNR is simulated by giving a velocity to the SNR and the surrounding ISM.

\begin{figure}
\center
\epsscale{1.15} \plotone{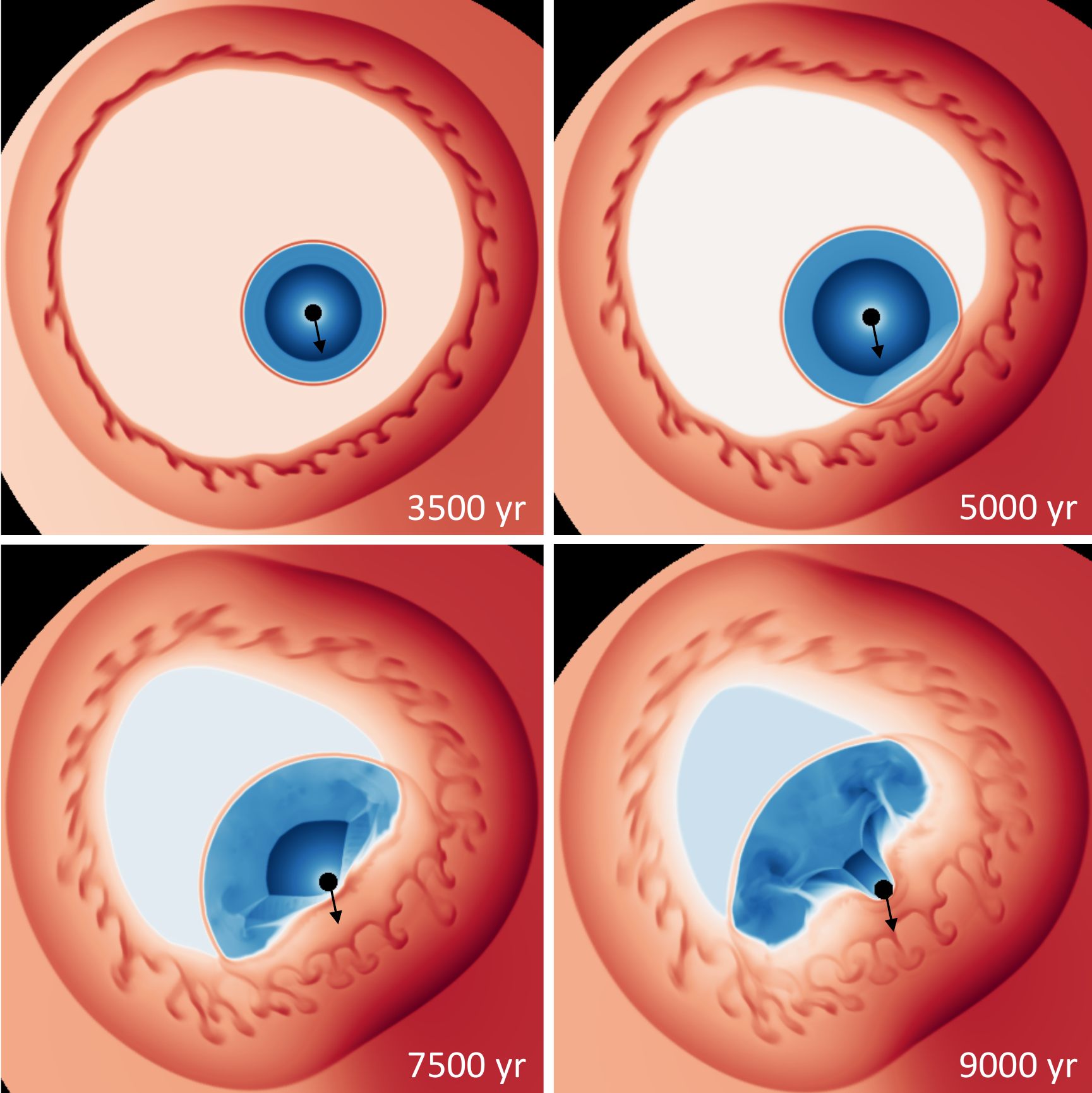}\caption{\label{fig3} Density maps at four different SNR ages corresponding to the HD simulation with input parameters summarized in Table~\ref{tab2}. The density scale is the same as in the left panel of Figure~\ref{fig2}. The direction of pulsar's motion is indicated by the black arrow. The average radius of the SNR shell in the four panels is 10~pc, 13~pc, 17~pc, and 19~pc, in order of increasing SNR age. }
\end{figure}

\subsection{Model Parameters}\label{param}

The HD model input parameters are summarized in Table~\ref{tab2}. The primary observational constraint included in the model is the pulsar's speed, measured in Section~\ref{proper} to be 720 $\rm km\:s^{-1}$. The SN explosion energy was set to $\rm 10^{51}\:erg$, and the other input parameters varied such that the morphology of the PWN and the SNR resembles the observed morphology at the time when the SNR radius reaches a size of $\sim$~21~pc (the observed radius of 18\arcmin\ at a distance of 4.1~kpc). The pulsar's initial spin-down luminosity $\dot{E}_0$ and and timescale $\tau_0$ were chosen to obtain a current spin-down luminosity $\dot{E}$ that is consistent with the total X-ray luminosity of the PWN measured in \citet{temim13a}. In this case, the ratio $L_X/\dot{E}=0.00235$ is within the range of luminosities observed for other pulsars \citep{possenti02}. The mass of the SN ejecta was set to 10~$\rm M_{\odot}$, the low end value predicted by the models of \citet{sukhbold16}. The minimum ISM density was set to $n_0=0.1\: \rm cm^{-3}$ \citep{temim13a}, increasing by a factor of eight in a direction perpendicular to the direction of the pulsar's motion.

\subsection{Results}

The simulation produced from the input parameters described in Section~\ref{param} is shown in Figure~\ref{fig2}. The left panel is a density map at an SNR age of 11,000 years, which is when the SNR reaches a radius of 21~pc and the SNR morphology most closely resembles the observed radio emission from the PWN shown in the left panel of Figure~\ref{fig1}. An increase in the ambient density would increase the SNR age, but would also cause the reverse shock to interact with the PWN sooner, resulting in a fully disrupted nebula at a time when the SNR shell reaches the observed radius. The early reverse shock interaction can be mitigated by increasing the SN ejecta mass. A combination of higher ambient density and higher SN ejecta mass can therefore produce a similar morphology as seen in Figure~\ref{fig2} at an age greater than 11,000 years.  

Using the SNR age of 11,000 years, a distance of 4.1 kpc, and the measured velocity of the pulsar, we can estimate the location of the SN explosion center, marked by the black cross in  Figure~\ref{fig1}~(left). The error box for the explosion center, reflecting the uncertainties in the pulsar's velocity and direction of  motion, is also shown. As expected, the site of the explosion is displaced to the west of the geometric center of the SNR shell, towards the direction of the higher ambient density. The radio morphology of the SNR shell also shows evidence that the radius of the shell is smaller in the western half of the SNR. The higher ambient density in this direction and the displacement of the PWN to the south that results from the pulsar's motion cause the SN reverse shock to interact with the PWN asymmetrically. This can be seen in Figure~\ref{fig3}, which shows four different time snapshots of the SNRs evolution. The PWN becomes displaced from the explosion center due to the pulsar's motion, and the reverse shock begins interacting with the nebula at $\sim$ 5000 years. At the age of $\sim$ 7500 years, the reverse shock reaches the pulsar, and as it continues towards the northeast, it sweeps the bulk of the PWN material away from the pulsar. A trail of PWN material begins to form  behind the pulsar, along the direction of the passage of the reverse shock (9000~yr panel of Figure~\ref{fig3}). This supports the finding of \citet{kolb17} that the PWN trail reflects the velocity flow of the SN ejecta. Due to the lower ambient density to the east, the SN reverse shock has not yet reheated all of the SN ejecta in the northeastern part of the SNR and has not yet reached the PWN from this direction. The unshocked ejecta material can be seen in the light blue color in Figure~\ref{fig2}~(left). The density enhancement at the northeastern edge of the PWN represents the innermost SN ejecta that have been swept up by the PWN.

The right panel of Figure~\ref{fig2} shows the map of the age of the particles ejected by the pulsar at the final SNR age of 11,000 years. While we do not take into account the effects of any enhanced synchrotron cooling that may be caused by the PWN's interaction with the reverse shock, the map is consistent with the observed radio and X-ray morphologies of the SNR, assuming that the particles cool with age. As can be seen in Figure~\ref{fig1}, the X-ray emitting region of the PWN in MSH 15-5\textit{6} coincides with the youngest particles concentrated in the trail behind the pulsar, shown in blue. The particles with an age greater than 4000 years shown in orange are concentrated in the northeastern region of the PWN that coincides with the PWN's radio emission, as would be expected from synchrotron aging.  The simulation also shows filamentary structures of younger particles that extend towards the northeast. These structures may explain the hard X-ray arc seen in the \textit{Chandra} image in Figure~\ref{fig1} (right), but since the arc traces the edge of the radio PWN and the location of the reverse shock, it is more likely that it arises from a jet that has been swept back by the shock, similar to the structure inferred for the Geminga pulsar \citep{posselt17}.

\section{Conclusion}\label{conclusion}

In this paper, we measured the proper motion of the pulsar in SNR~MSH~15-5\textit{6} using \textit{Chandra} X-ray images taken at two different epochs, resulting in a baseline of $\sim$ 15~years. We measure a pulsar velocity of $720^{+290}_{-215}\rm km\: s^{-1}$ and a direction of motion of $14^{\circ}\pm20^{\circ}$ west of south. 
We used the measured pulsar velocity to constrain an HD model of the evolution of the SNR that invokes a moving pulsar and a density gradient in the surrounding ISM that cause the reverse shock to interact with the PWN asymmetrically. The model reproduces the morphology of the SNR at an age of 11,000 years, assuming an explosion energy of $10^{51}\rm\: erg$, an SN ejecta mass of 10~$\rm M_{\odot}$, an average ambient density of $0.4\rm \: cm^{-3}$, and a distance of 4.1~kpc. The measured pulsar velocity places a constraint on the orientation of the ambient density gradient and indicates that the density increases from east to west. An increase in the average ambient density and the SN ejecta mass would produce a similar SNR morphology at an older SNR age. Constraints on the ambient density from upcoming  \textit{XMM-Newton} observations of the entire SNR shell will allow us to use the HD model to better constrain the range of SN ejecta masses and ages that could produce the observed morphology of the SNR.

\acknowledgments

Support for this work was provided by the National Aeronautics and Space Administration through Chandra Award Number GO6-17057X issued by the Chandra X-ray Center, which is operated by the Smithsonian Astrophysical Observatory for and on behalf of the National Aeronautics Space Administration under contract NAS8-03060.

\bibliographystyle{apj}

\end{document}